\documentclass[11pt,preprint]{aastex}

\newcommand{\water}{${\rm H}_{2}{\rm O}$ }

\newcommand{\pp}{\prime \prime}

\newcommand{\methanol}{{$\rm{CH}_{3}{\rm OH}$}}

\shorttitle{Performance of KVN radio telescopes}
\shortauthors{Lee et al.}

\begin{document}


\title{Single dish performance of KVN 21-m radio telescopes :\\
       Simultaneous observations at 22 and 43\,GHz}


\author{Sang-Sung Lee\altaffilmark{1,2},
Do-Young Byun\altaffilmark{1,2},
Chung Sik Oh\altaffilmark{1,2},
Seog-Tae Han\altaffilmark{1,2},
Do-Heung Je\altaffilmark{1},
Kee-Tae Kim\altaffilmark{1,2},
Seog-Oh Wi\altaffilmark{1},
Se-Hyung Cho\altaffilmark{1,3},
Bong Won Sohn\altaffilmark{1,2},
Jaeheon Kim\altaffilmark{1,2,4},
Jeewon Lee\altaffilmark{1,4},
Se-Jin Oh\altaffilmark{1},
Min-Gyu Song\altaffilmark{1},
Jiman Kang\altaffilmark{1},
Moon-Hee Chung\altaffilmark{1},
Jeong Ae Lee\altaffilmark{1,5},
Junghwan Oh\altaffilmark{1,2},
Jae-Han Bae\altaffilmark{1},
So-Young Yun\altaffilmark{1},
Jung-Won Lee\altaffilmark{1},
Bong Gyu Kim\altaffilmark{1,2},
Hyunsoo Chung\altaffilmark{1},
Duk-Gyoo Roh\altaffilmark{1},
Chang Hoon Lee\altaffilmark{1},
Hyun Goo Kim\altaffilmark{1},
Hyo Ryoung Kim\altaffilmark{1},
Jae-Hwan Yeom\altaffilmark{1},
Tomoharu Kurayama\altaffilmark{1},
Taehyun Jung\altaffilmark{1},
Pulun Park\altaffilmark{1,2},
Min Joong Kim\altaffilmark{1,6},
Dong-Hwan Yoon\altaffilmark{1},
and
Won-Ju Kim\altaffilmark{1,7}
}
\altaffiltext{1}{Korean VLBI Network,
Korea Astronomy and Space Science Institute, 
P. O. Box 88, Yonsei University, Seongsan-ro 262, Seodaemun, Seoul 120-749, 
Republic of Korea;sslee@kasi.re.kr}

\altaffiltext{2}{Yonsei University Observatory, Yonsei University,
Seongsan-ro 262, Seodaemun, Seoul 120-749, Republic of Korea}

\altaffiltext{3}{Department of Astronomy, Yonsei University, Seongsan-ro 262,
Seodaemun, Seoul 120-749, Republic of Korea}

\altaffiltext{4}{Department of Astronomy and Space Science,
Kyung Hee University, Seocheon-Dong, Giheung-Gu, Yongin, Gyeonggi-Do, 446-701}

\altaffiltext{5}{University of Science and Technology, 133 Gwahangno,
Yuseong-gu, Daejeon, 305-333, Republic of Korea}

\altaffiltext{6}{Department of Astronomy and Space Science,
Sejong University, Seoul 143-747, Korea}

\altaffiltext{7}{Department of Astronomy and Space Science,
Chungnam National University, Daejeon 305-764, Korea}

\begin{abstract}
We report simultaneous multi-frequency observing performance 
at 22 and 43\,GHz of 
the 21-m shaped-Cassegrain radio telescopes of the Korean VLBI Network (KVN).
KVN is the first millimeter-dedicated VLBI network in Korea 
having a maximum baseline length of 480\,km.
It currently operates at 22 and 43\,GHz 
and planed to operate in four frequency bands, 22, 43, 86, and 129\,GHz. 
The unique quasioptics of KVN enable simultaneous multi-frequency observations
based on efficient beam filtering and accuarate
antenna-beam alignment at 22 and 43\,GHz. 
We found that the offset of the beams is within $<5$ arcseconds
over all pointing directions of antenna.
The dual polarization, cooled HEMT receivers at 22 and 43\,GHz result in 
receiver noise temperatures less than 40\,K at 21.25-23.25\,GHz and 
80\,K at 42.11-44.11\,GHz.
The pointing accuracies have been measured to be 3 arcseconds in azimuth and 
elevation for all antennas. 
The measured aperture efficiencies are 
$65\%$(K)/$67\%$(Q), 
$62\%$(K)/$59\%$(Q), 
and $66\%$(K)/$60\%$(Q) 
for the three KVN antennas, KVNYS, KVNUS, and KVNTN, respectively.
The main-beam efficiencies are measured to be 
$50\%$(K)/$52\%$(Q), 
$48\%$(K)/$50\%$(Q), 
and $50\%$(K)/$47\%$(Q) 
for KVNYS, KVNUS, and KVNTN, respectively.
The estimated Moon efficiencies are 
$77\%$(K)/$90\%$(Q), 
$74\%$(K)/$79\%$(Q),
and $80\%$(K)/$86\%$(Q) 
for KVNYS, KVNUS, KVNTN, respectively.
The elevation dependence of the aperture efficiencies is quite flat for
elevations $> 20^{\circ}$.

\end{abstract}


\section{Introduction}

The Korean VLBI Network (KVN) is the first mm-dedicated 
Very Long Baseline Interferometry (VLBI) network 
in East Asia, aiming to study the formation and death
of stars, the structure and dynamics of our own Galaxy,
and the nature of active galactic nuclei (AGNs)~\citep{kim+04,waj+05}. 
Simultaneous multi-frequency VLBI observations 
using the multi-frequency band receiver systems~\citep{han+08}
will enable us to conduct observational studies with
high spatial and temporal resolution based on 
the technique of
multi-frequency phase referencing~\citep{asa+98,mid+05,jun+11}.

The KVN project was started in 2001 by 
the Korea Astronomy and Space Science Institute (KASI)
with the construction of three 21-m radio telescopes in Seoul, 
Ulsan and Jeju island, Korea; 
KVN Yonsei Radio Telescope (KVNYS),
KVN Ulsan Radio Telescope (KVNUS),
and KVN Tanma Radio Telescope (KVNTN).
The telescopes are located 
on the campuses of Yonsei University (Seoul), 
Ulsan University (Ulsan),
and Tamna University (Jeju island).
KVN can operate simultaneously in two radio frequency bands, 
22 and 43\,GHz,
and will be upgraded for simultaneous four-frequency observations
including the higher frequencies at 86 and 129\,GHz.
The maximum baseline length is $\sim$ 480\,km and the maximum angular
resolution at the highest operating frequency (129\,GHz) is 
$\sim$ 1\,milliarcsecond (mas).

A single dish first light was received by the KVN Yonsei Radio Telescope 
on 30 August, 2008. 
Following the single-dish performance evaluation observations,
all three radio telescopes of KVN are now operational,
and actively operated for single-dish observations of
\water, SiO, and Methanol maser sources and extragalactic compact radio
sources~\citep{kim+10}. 
In this paper we introduce the general aspects
of the radio telescopes of the new mm-VLBI network and describe 
the results of the performance evaluation observations. 
In section 2, 
we describe general aspects of observing systems of the Korean VLBI Network.
We report the results of the performance evaluation observations 
for the single dish radio telescopes in Section 3. 
In Section 4, 
some astronomical results using KVN radio telescopes are described.
Our results are summarized in Section 5.

\section{Telescope system}
The KVN project aimed to build the first 
millimeter-dedicated VLBI network in Korea. 
KVN has a unique observing system,
allowing simultaneous multi-frequency single polarization oberving
in up to four frequency bands: 22, 43, 86, and 129\,GHz 
or simultaneous multi-frequency dual polarization observing
in two frequency bands out of four.
A new millimeter-wave receiver optics system with three frequency-selective
surfaces (dichroic filters) is an essential part of the unique system
\citep{han+08}.
In this section we describe the main technical features of the KVN antennas
and the observing system. 

\subsection{Antennas}
\begin{figure}
\epsscale{1.0}
\plotone{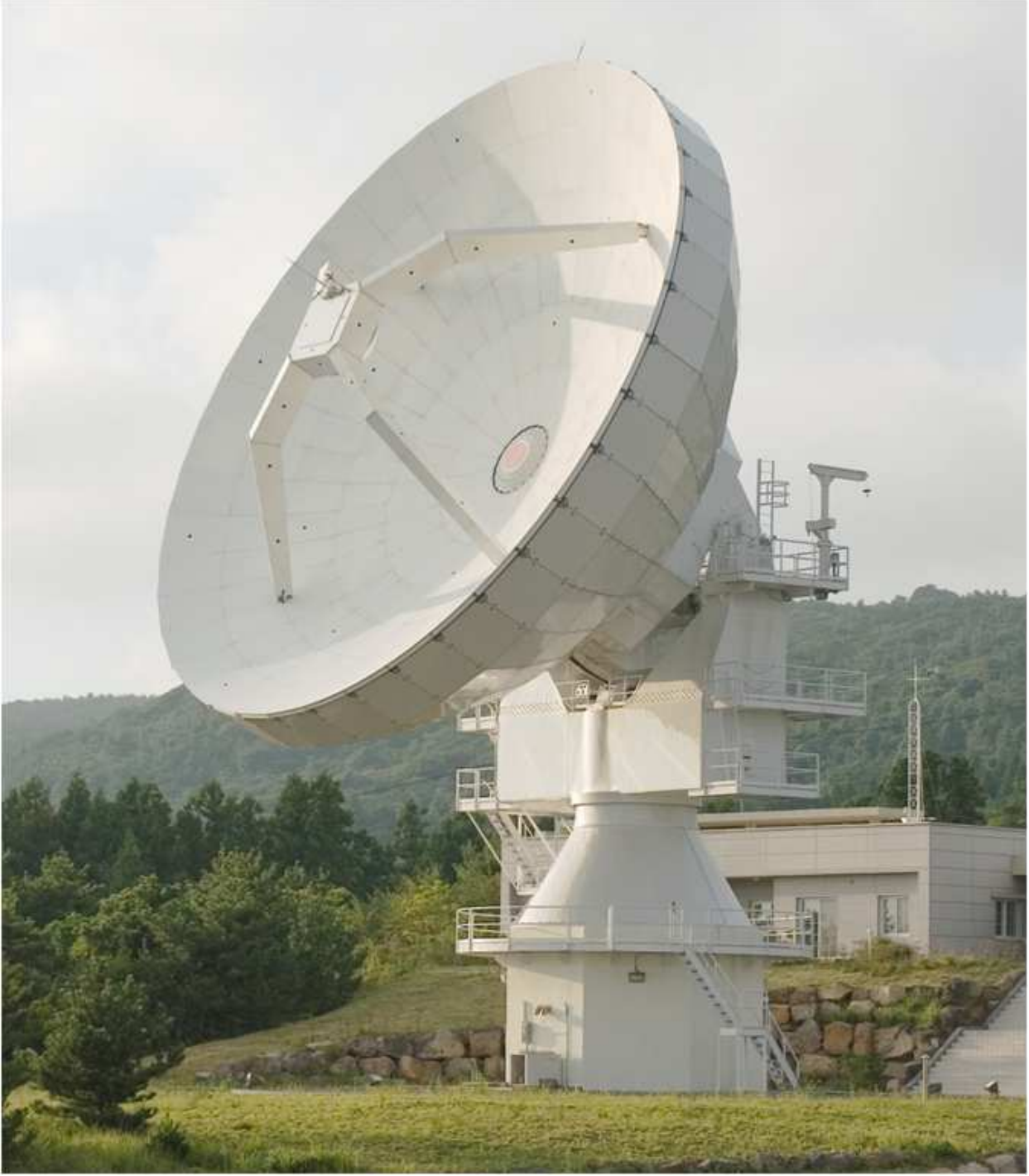}
\caption{Photograph of the 21-m diameter KVN Tamna radio telescope. 
\label{fig-ant}}
\end{figure}

The three 21-m antennas of KVN have been 
constructed by KASI under a contract with
an American company, Antedo, Inc., and a Korean company, HighGain 
from 2004 to 2008. The antennas are located at University campuses which
have the well-developed infrastructures and 
yield the optimal VLBI baselines for
the Korean VLBI Network in the Korean pennisula.
The main design features of the reflectors are determined by the fact 
that the antennas manintain a high surface accuracy of $<150\mu$m RMS 
and a pointing accuracy of $<$4 arcsecond at wind speeds $<$10m/sec,
and the aperture efficiency should be high enough
(e.g., $\sim$60\% at 100\,GHz) as a mm-dedicated
VLBI network. The KVN antennas are therefore designed to be 
a shaped Cassegrain-type antenna with an alt-az mount (Figure~\ref{fig-ant}). 
The telescope has a 21-m diameter main reflector with 
a focal length of 6.78\,m. 
The main reflector consists of 200 aluminium panels with a manufacturing
surface accuracy of $\sim 65 \mu$m. 
The slewing speed of the main reflector is 
$3^{\circ}\,\rm{sec}^{-1}$, which enables 
fast position-switching observations.
The subreflector position, tilt, and tip
are remotely controlled and modeled to compensate for 
the gravitaional deformation of the main reflector
and for the sagging-down of the subreflector itself.
The characteristics of the antenna systems are summarized in 
Table~\ref{tbl-1}.

\subsection{Quasi optics}

A millimeter-wave receiver optics system with three frequency-selective
surfaces (dichroic filters) is an essential component of the unique system
enabling simultaneous multi-frequency VLBI observations 
in four frequency bands, 22, 43, 86 and 129\,GHz. 
The new receiver optics system has been desribed by \cite{han+08}, 
and only its salient features will be described in this section. 

\begin{figure}
\epsscale{1.0}
\plotone{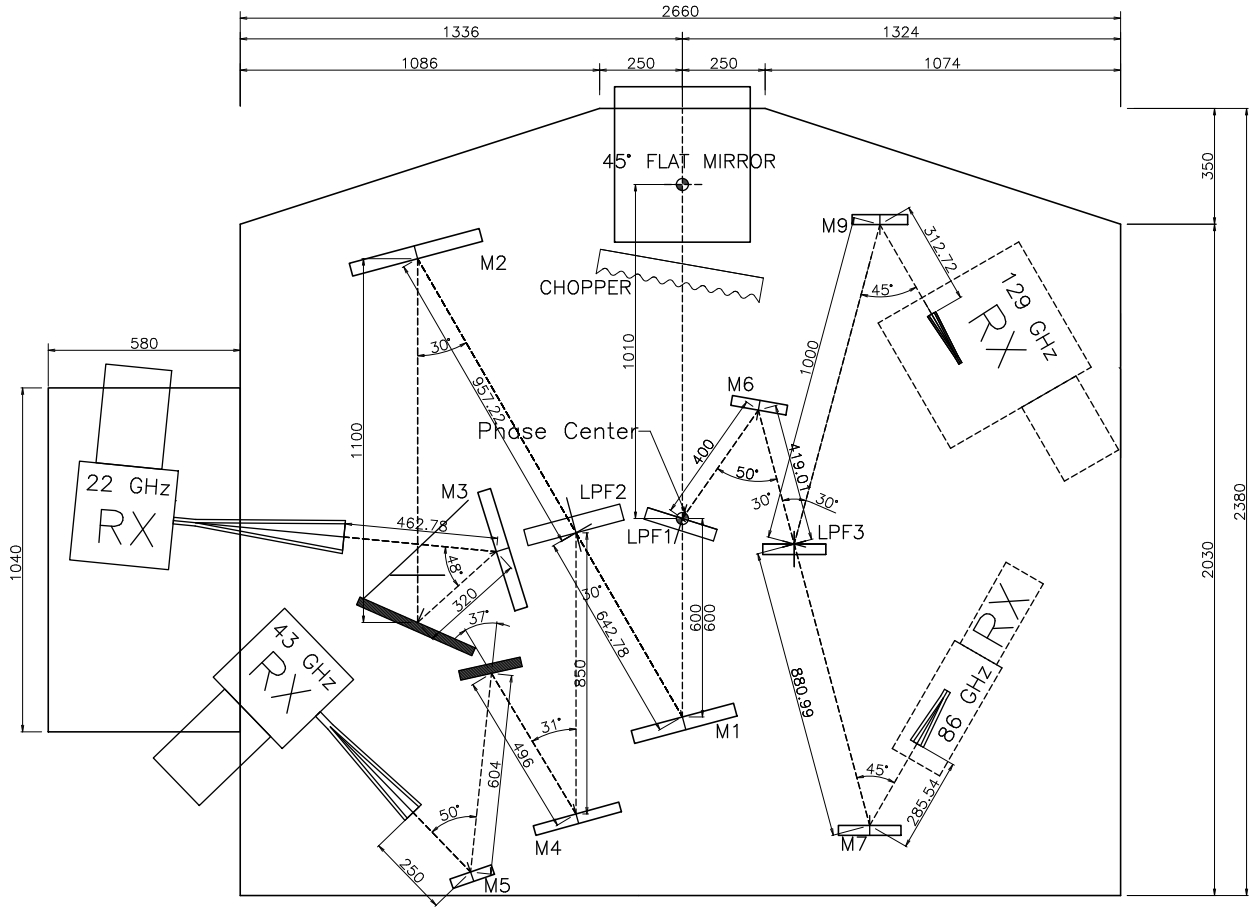}
\caption{Layout of KVN receiver optics; top view, 
the beam from the antenna comes downwards. 
The 86 and 129\,GHz receivers and their corresponding 
quasioptical systems (M6, M7, M8, and LPF3) are currently being installed. 
Dimensions are given in millimeters, and angles are in degrees. 
}
\label{layout}
\end{figure}

In order to make simultaneous observations in the four frequency bands, 
receiver optics system was deveploped 
to split one signal into four using 
three low pass filters with nominal cutoffs 
at 30, 70, and 108\,GHz, made of multi-layer metal meshes. 
The beam passing through the filter must have a large size. 
Furthermore since each band is to provide dual polarization observation 
the receiver optics should generate as little cross-polarization as possible.
As usual with shaped Cassegrain antenna optics,
the edge illumination at subreflector by the receiver optics
should be about 17dB i.e. 
less than 12dB in conventional Cassegrain antenna optics, 
which leads to 2\% spillover loss, resulting in higher illumination 
efficiency than obtainable from conventional Cassegrain optics 
fed with a scalar horn. 
Satisfying such requirements usually conflicts with physical space 
available inside an antenna cabin. The layout of the complete 
receiver optics is shown in Figure~\ref{layout}. 
This configuration was determined iteratively using standard 
quasioptical theory~\citep{gol98} rather than computationally 
intensive physical optics method. 
The beam from the subreflector comes downwards 
to the top of the 45$^{\circ}$ flat mirror. 
A microwave absorber may be inserted between
the flat mirror and the first Low Pass Filter
for system calibration based on the chopper-wheel method~\citep{uh76}.
The beam from the flat mirror 
may be reflected, pass through, or be filtered by three Mode Selectors
which consist of a flat mirror, a hole (or free space), 
and a Low Pass Filter (LPF).
LPF1 (dichroic, cutoff$\sim$70\,GHz) reflects 
the beam for the 86/129\,GHz branches
and lets the beam for the 22/43\,GHz branches pass through. 
Likewise LPF2 (cutoff$\sim$30\,GHz) and LPF3 (cutoff$\sim$108\,GHz) 
split beams into 22/43\,GHz and 86/129\,GHz respectively. 
The 86 and 129\,GHz receivers and their corresponding quasioptical systems
are currently being installed at three telescopes.
Corrugated horns for 86 and 129\,GHz are 
to be located inside cryogenic dewars 
while horns for 22 and 43\,GHz reside outside the respective dewars.
The signal losses due to the dichroic filters
were measured to be $<2\%$(LPF1), $<4\%$(LPF2), $<5\%$(LPF1+LPF2)
for the 22\,GHz-band
and  $<2\%$(LPF1), $<6\%$(LPF2), $<7\%$(LPF1+LPF2)
for the 43\,GHz-band~(Table \ref{tbl-lpf}).

\subsection{Receivers}

The 22 and 43\,GHz band signals separated in the quasioptical system
are coupled into the corrugated feed horns
and receivers which have been built
at KASI. 
The observing frequency bands for the receivers are 
21.25-23.25\,GHz and 42.11-44.11\,GHz, respectively. 
Both receivers are dual circular polarization, 
cooled HEMT receivers.

\begin{figure}
\epsscale{1.0}
\plotone{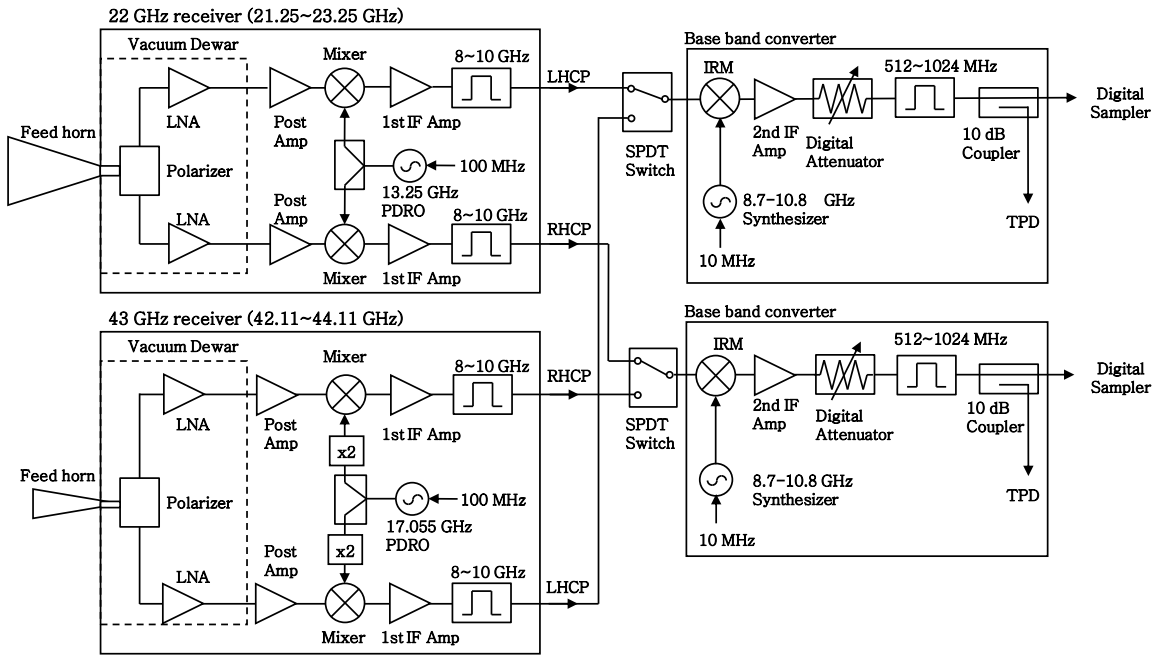}
\caption{Simplified schematic diagram of 22 and 43\,GHz receivers;
The feed horn input signals are transmitted to the digital sampler
and total power detector (TPD) through LNA (Low Noise Amplifier),
following a low noise amplifier (LNA), additional amplifications,
two frequency conversions, and signal leveling. 
Two signals are selected using two SPDT switches.
PDRO: Phase-locked Dielectric Resonator Oscillator,
x2: frequency doubler,
LHCP: Left Handed Circular Polarization,
RHCP: Right Handed Circular Polarization,
SPDT: Single Pole Double Throw,
IRM: Image Rejection Mixer,
TPD: Total Power Detector.
}
\label{fig:rx}
\end{figure}

As shown in Figure~\ref{fig:rx}, 
LHCP (Left Handed Circular Polarization) and 
RHCP (Right Handed Circular Polarization) components of the 
signals are separated in septum polarizers  
inside the dewar of each receiver. 
The InP HEMT amplifiers made by Caltech and NRAO are used as the first stage 
LNA (Low Noise Amplifier) for the 22 and 43\,GHz receivers. 
The equivalent noise temperatures of both amplifiers are 
approximately 15\,K over the whole observing frequency bands. 
The receiver noise temperatures for three 22\,GHz receivers
are measured to be 30-40\,K. For the 43\,GHz receivers,
the receiver noise temperatures are 40-50\,K for one receiver 
and 70-80\,K for the other two receivers. 
The increase of the receiver noise temperatures is resulted from
the aggregated loss by feed horn, vacuum window, thermal isolator,
and polarizer.
The 43\,GHz receiver with lower noise temperature is thermally isolated
by a thermal isolator adopted at VLA, whereas the others by
photonic bandgap isolators between dewar and polarizer.  
The spread of the receiver noise temperatures among 43\,GHz receivers
may be due to the difference of the thermal isolators.
The signals are first amplified in a cooled stage
and then further amplified by a room-temperature amplifier.
They are then down-converted to 8-10\,GHz, 
$1^{\rm st}$ IF (Intermediate Frequency) in the mixers using PDROs
(Phase-locked Dielectric Resonator Oscillator) which 
are locked on the 100\,MHz reference signal transmitted from the H-maser. 
The observing frequency bandwidth is limited within 2\,GHz 
by BPF (Band Pass Filter) in $1^{\rm st}$ IF. 
Two $1^{\rm st}$ IF signals are selected using SPDT switches.
The selected signals may consist of single polarization signals at dual frequency bands or dual polarization signals at single frequency band.
The signal conditioning and $2^{\rm nd}$ frequency conversion 
are done in the base band converter 
by using IRM (Image Rejection Mixer)
with an image rejection ratio of 18\,dB, 
a 8.7-10.8\,GHz frequency synthesizer with a step frequency of 100\,Hz, 
and a digital step attenuator which offers an attenuation range
up to 31.5\,dB in steps of 0.5\,dB. 
The final analog output signals are limited to
512-1024\,MHz by an anti-aliasing filter.

\subsection{Backends}
The output signal from the  base band converter is split into two different paths.
One goes to a digital sampler and 
the other to a square-law detector (or TPD).
The output from the detector is amplified and input into a voltage-to-frequency converter (hereafter VFC).
An embedded counter of the VFC counts the output pulse rates, which are propotional to the output power of 
the base band converter. 
The counted output pulse rates are taken every 100 millisecond by a receiver-control computer in the receiver cabin.
The VFC data are used
not only for calibrating the receiver gain and the sky attenuation level 
but also for measuring the total flux of celestial objects in continuum observation mode. 
An observing technique, cross scan, is used for the total flux measurements
of point-like objects in continuum single dish observations at KVN.
Since the technique relies on scanning with the main beam over the source
position in both azimuth and elevation directions,
the receiver gain and atmosphere fluctuations faster than
a few seconds could not be tracked and removed.
The calibration against the faster fluctuations will be improved
by using a noise diode with known temperature
and a faster data aquisition system. 

\begin{figure}
\epsscale{1.0}
\plotone{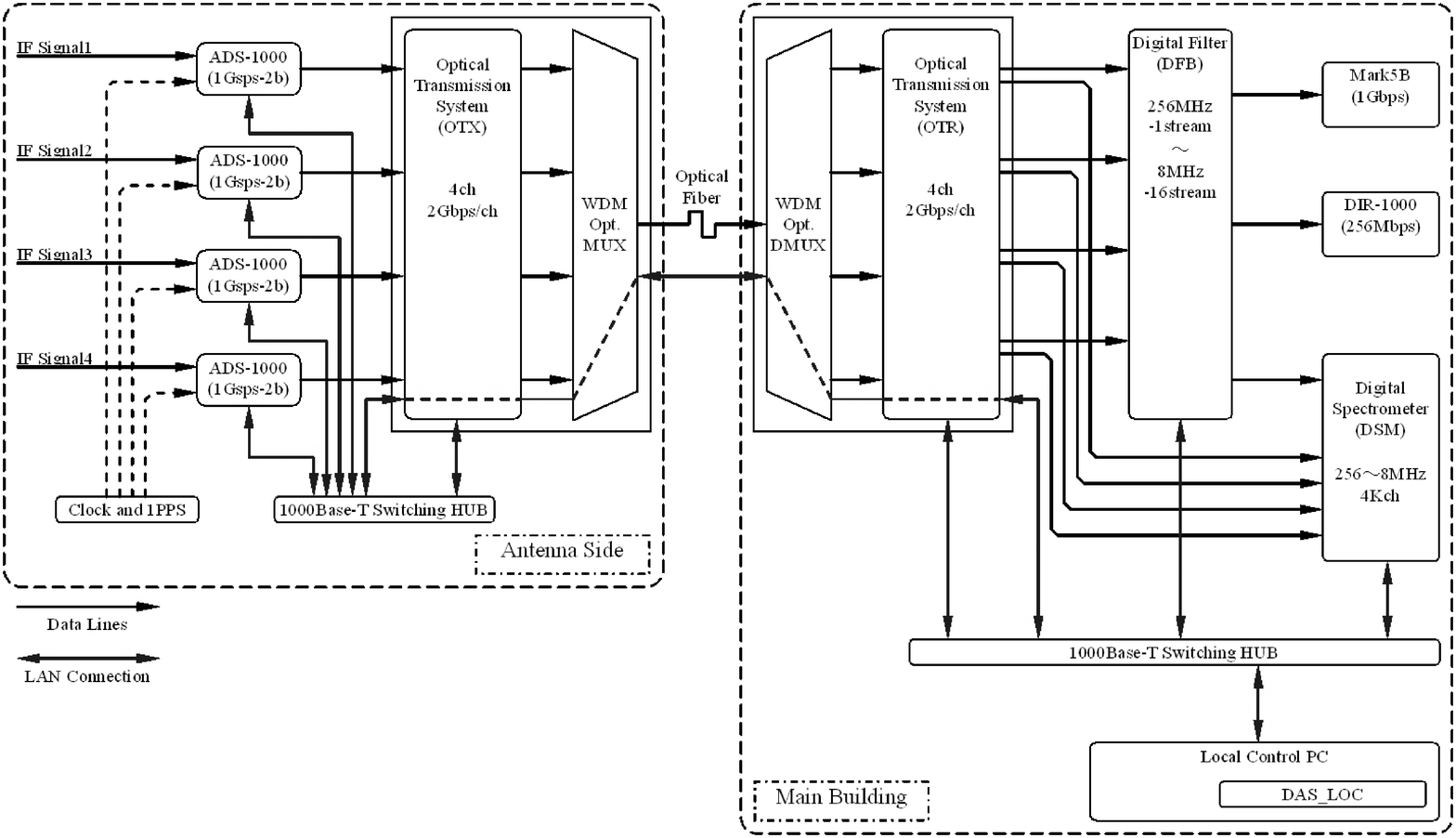}
\caption{Schematic diagram of KVN Data Acqusition System.
Adopted from \cite{oh+11}
\label{fig:das}}
\end{figure}

The signals digitized by the samplers in the receiver room are processed 
by the KVN Data Acqusition System (DAS) to get spectra for single-dish 
spectroscopy observations.
In VLBI operation, the digitized signals processed in the DAS are recorded onto storage media such as harddisks or magnetic tapes.
As shown in Figure~\ref{fig:das}, the DAS consists of four 
subsystems; 
samplers, optical trasmission system (OTS), digital filter bank (DFB) and digital spectrometer (DSM),
which were manufactured by 
a Japanese company, Elecs Industry Co. Ltd.
The samplers digitize base band signals into 2-bit data streams with four quantization levels. The output data streams of the samplers are 
transmitted to the observing building 
through optical fibers by the OTS.
The DAS is configurable to various modes according to the required number of streams and bandwidths.
For wideband observations, the DSM uses the output streams of the OTS, 
while for narrow band observations it uses the output stream of the DFB.
In principle, the DFB produces 16 data streams of 16\,MHz bandwith from 4 streams of 512\,MHz bandwidth.
Combining more than one stream, the DFB can produce streams with wider bandwidth such as 8$\times$32\,MHz, 4$\times$64\,MHz, 2$\times$128\,MHz and 1$\times$256\,MHz.
The DSM is a FX-type digital correlator and we use the auto-correlation output for spectroscopy observations.
Table~\ref{table:obsmode} summarizes the available spectrometer outputs of the DSM.
The correlation output produced by the DSM has 8192 channels but control computer can take only half of them.
So we take full bandwidth after smoothing and binning
pairs of adjacent channels,
or half bandwidth by selecting 4096 contiguous channels.
For example in the wide mode, we can get 4$\times$512\,MHz bandwidth spectra or 4$\times$256\,MHz bandwidth spectra.
Detailed information for the KVN DAS is described in \cite{oh+11}.
The DSM can calculate the cross-correlation of two input data streams
and thus can be used for polarimetry.
The polarimetry observation mode using KVN single-dish is currently being test,
and will be described in a separate paper.

\subsection{Observation control software}
For single-dish operation using a KVN antenna, 
we support various observation modes such as sky dipping (SDIP), 
position switching (PS), 
frequency switching (FS), focusing (FOCUS), 
five-point mapping (FIVE), grid mapping (GRID), 
cross scan (CS) and on-the-fly mapping (OTF).
The observed data from all observing modes except for the continuum OTF mapping are stored into a Gildas CLASS file.
The continuum OTF data written in ASCII format are reduced and converted 
into FITS format.
The observation control software was written in Python progmaming language 
and runs on IPython interpreter.
A sequence of observations for many sources can be easily run by writing 
a simple Python script.
The observation software runs on a PC running Linux.
At most 8 streams of data can be simultaneously processed with a data-taking
cycle of 100 milliseconds.
A graphical user interface (GUI) is used for configuring devices 
and running observations.

\section{Performance tests}

On 10th October, 2008, we had first light with the KVN
Yonsei radio telescope 
simultaneously observing \water and SiO masers in Orion KL at 22 and 43\,GHz.
The resultant spectra at both frequency bands indicate 
that the overall receiving system of KVN Yonsei radio telescope
is functioning properly. In this section, we describe 
the results of single dish performance tests 
for KVN radio telescopes.

\subsection{Antenna pointing performance}
\subsubsection{Beam alignment}

It is very important to 
align the beams of different bands with other as accurately as possible. 
Poor beam alignment will result in the large 
uncertainty of flux measurements
for simultaneous multi-frequency obsevations.
Therefore the beam alignment has to be measured
and corrected to satisfy the requirement
that two (or four) beams should be aligned within $<10\%$ of the FWHM of 
the smallest beam 
(e.g., 7 arcseconds for 22 and 43\,GHz, 
and 2.5 arcseconds for 22, 43, 86, and 129\,GHz).

As a first step of the beam alignment measurements,
we observed some bright and compact radio sources at 22 and 43\,GHz 
including planets and SiO maser sources for measuring the beam alignment
after the setup of the quasioptics.
The beam alignments were estimated by measuring the pointing offsets
at each frequency with CS and FIVE observing modes.
If the measured beam alignments were larger than the requirement,
we have investigated the reasons for the poor alignment. 
For all cases, the poor beam alignments were caused 
by the misalignment of the quasioptical system including feed horns, 
and could be corrected by changing the angle of flat mirrors 
and feed horns in the quasioptical system.
Table~\ref{tab-bl} shows that the results of beam alignments 
with respect to the beam at 43GHz for 
three KVN telescopes after corretions.
The beams of all KVN antennas were very well aligned in azimuth
and there was no need for correction. However, beam offsets
in elevation were measured to be as large as 7 arcseconds. 
After changing the angle of
flat mirrors in the quasioptical system, we measured the beam offsets
to be $<5$ arcseconds. The elevation and azimuth dependence of the beam alignments
also have been investigated by observing planets and SiO masers at
various elevations. 
We found no significant elevation dependence of the beam alignments for all KVN antennas.
This implies that the quasioptical systems and the receiver plates including
the sub-plate for the 22\,GHz receivers are stable against gravitational
deformation.

\subsubsection{Pointing accuracy}

Since the beams at 22 and 43\,GHz are well aligned each other,
it is efficient to establish the pointing model only at 43\,GHz using 
SiO maser sources (mostly late type stars) whose positions are very well
known. As long as the alignment between two antenna beams
is accurate to $<10\%$ of the FWHM at 43\,GHz,
the pointing model derived from SiO maser observations at 43\,GHz
should ensure the pointing accuracies for both frequency bands. 

\begin{figure}
\epsscale{1.0}
\plotone{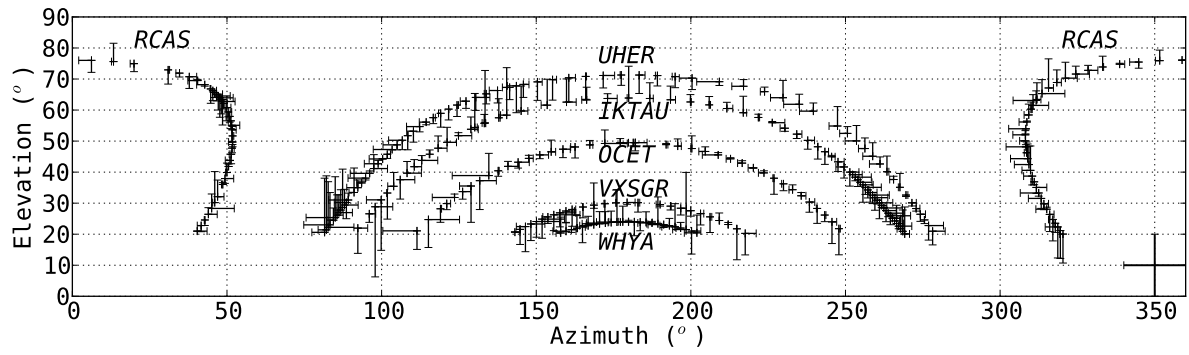}
\caption{Residual pointing offsets of the pointing model 
for KVNYS established on 16 October, 2010 through observations of
6 SiO maser sources. Cross in the lower right corner
indicates the scale of error bars at 10 arcseconds both
in azimuth and elevation.
\label{fig:res}}
\end{figure}

A pointing model had already been obtained by observing
strong SiO maser sources with a 100\,GHz receiving system.
These observations had been done to evaluate the performance
of the antenna under the contract with
the antenna manufacturer~\citep[see][]{kim+11}.
On the basis of this pointing model established at 100\,GHz,
a sample of 9 late type stars (Table~\ref{tab-poi1}) 
have been used continuously 
to improve the pointing accuracy of the telescope.
Table~\ref{tab-poi2} describes the accuracies of pointing models
established for three KVN radio telescopes 
during the evaluation period 2009-2011. 
The root mean square (rms) of the residual pointing offsets
between the observations and the pointing models are listed 
in azimuth ($\sigma_{\rm Az}$), elevation ($\sigma_{\rm El}$), and 
total ($\sigma_{\rm tot} = \sqrt{\sigma_{\rm Az}^2 + \sigma_{\rm El}^2}$),
respectively, for each epoch and telescope
(see Figure~\ref{fig:res} for an example of residual pointing offsets).
Before applying a cladding system (see below for detail), 
the residual rms errors are $<3$ arcsecond in azimuth and $<6$ arcsecond in elevation.
The pointing accuracies have been improved
to be $<3$ arcsecond in both azimuth and elevation 
after applying the cladding system 
for more efficiently insulating the antenna mount structure.

\begin{figure}
\epsscale{0.7}
\plotone{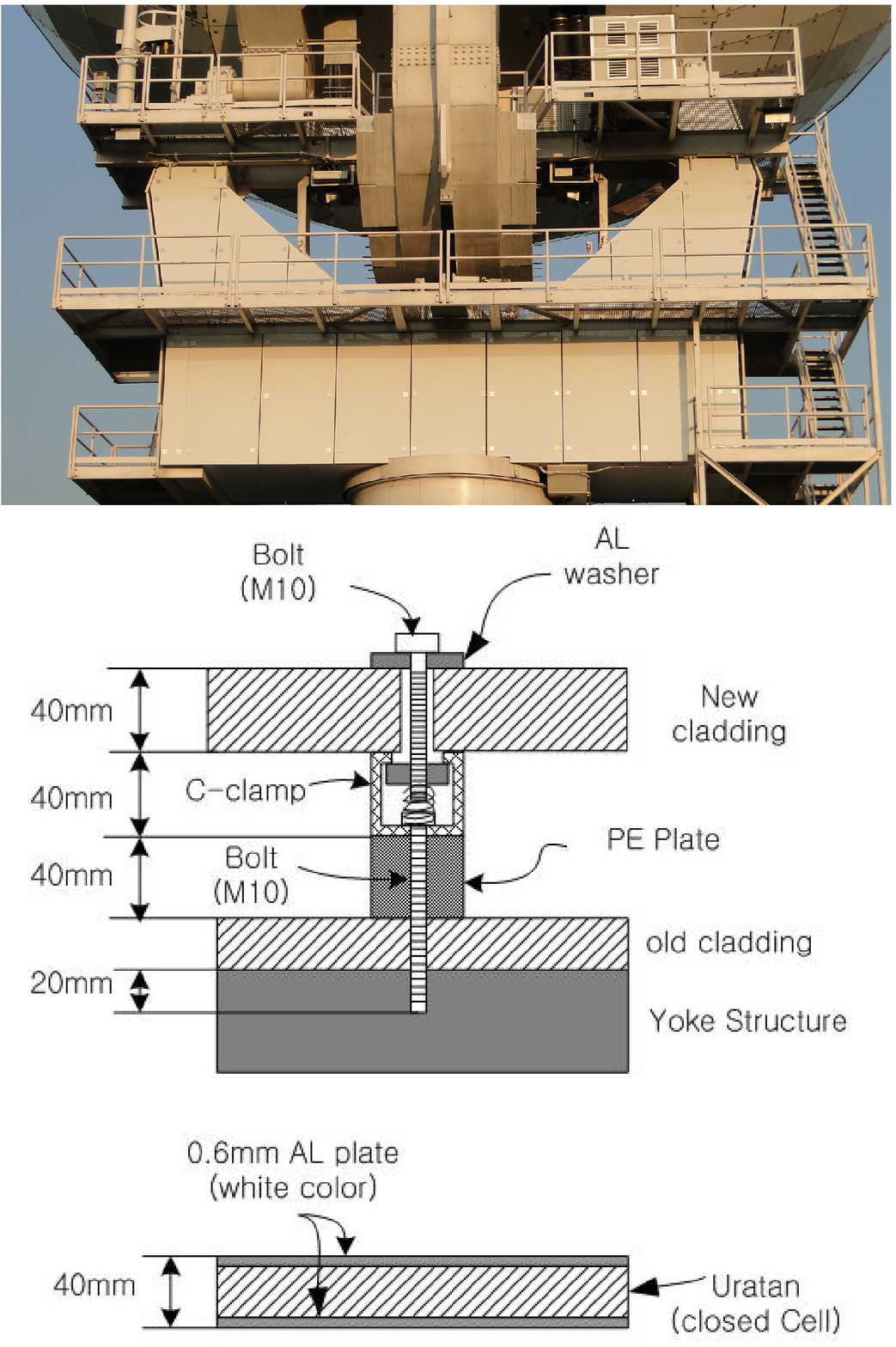}
\caption{Antenna mounting structure (Yoke arms) with new cladding system
(upper panel) and detailed structures of new cladding system (lower panel).
\label{fig:clad}}
\end{figure}

Thermal deflection of the antenna mounting structures (Yoke arms) have been 
investigated in order to imporve the antenna pointing accuracy.
When the antenna mounting structures face directly to the Sun,
thermal deflection causes the antenna pointing offsets to rapidly 
change from a few to $\sim$40 arcseconds within an hour.
The large pointing offsets result in a large uncertainty of flux measurements 
not only for single dish observations but also for VLBI (phase referencing)
observations during day time.
Therefore, we decided to apply a cladding system
to the antenna mounting structures (Figure~\ref{fig:clad}).
New cladding system provides an efficient thermal insulation
by making an air gap with its thickness of 80\,mm
between new and old cladding systems.
New cladding systems were installed 
on 14 October 2010, 6 November 2010, and 24 November 2010 for KVNYS, KVNUS,
and KVNTN, respectively.
After applying the cladding, 
the pointing offsets due to thermal effects have been reduced to be 
less than 20 arcseconds and
the time scale of the thermal effects improved to be $>$ 5 hours.
As a result, the rms residuals in the pointing model have been improved 
by a factor of $\sim 2$ (Table~\ref{tab-poi2}).
In case of observations during day time, especially at sunrise,
the pointing observations must be conducted out at least every hour
for maintaining the pointing accuracy to be less than 4 acrseconds.

\subsection{Antenna efficiency and gain curve}

\subsubsection{Antenna efficiency}

Surface accuracies of the main reflectors of the KVN radio telescopes
had been measured by photogrammetry at an elevation of $48^{\circ}$,
yielding alignment surface accuracies of 
50, 54, and 51\,$\mu m$ for the KVN Yonsei, Ulsan, and Tamna telescopes,
respectively~\citep{kim+11}. 
The expected total surface accuracy of KVN antennas is $\sim 124\,\mu$m
(Table~\ref{tbl-1})
which is good
to enable observations at frequencies of $< 150\,{\rm GHz}$.
The surface accuracies have been verified by the 100\,GHz 
test observations yielding aperture efficiencies of 52\% and
main-beam efficiencies of 46\% for all three telescopes~\citep{kim+11}.

Since new recievers and quasioptical systems at 22 and 43\,GHz have been
installed for KVN radio telescopes,
the corresponding antenna efficiencies 
should be measured in order to evaluate the alignment 
of the quasioptical systems including feed horns.
The antenna efficiencies can be
measured by observing primary calibrators such as planets.
Since a planet can be considered as disk with uniform brightness,
the measured antenna temperature, $T^{*}_{\rm A}$, is related to
the aperture ($\eta_{\rm A}$) and beam ($\eta_{\rm B}$) efficiencies as
~\citep{sch+80,rho+99,koo+03,kim+11}: 
\begin{equation}
\label{eqn:eff1}  
\eta_{\rm A} \equiv \frac{A_{\rm e}}{A_{\rm p}} =  
  \frac{\lambda^2 T^{*}_{\rm A}}{A_{\rm p}T_{\rm B}\Omega_{\rm s}},
\end{equation}
\begin{equation}
\label{eqn:eff2}  
\eta_{\rm B} \equiv \frac{\Omega_{\rm M}}{\Omega_{\rm A}}  =   
  \frac{T^{*}_{\rm A}\Omega_{\rm M}}{T_{\rm B}\Omega_{\rm s}} =
  \frac{A_{\rm p}\Omega_{\rm M}}{\lambda^{2}}\eta_{\rm A},
\end{equation}
\begin{equation}
\label{eqn:eff3}  
\Omega_{\rm s} = \Omega_{\rm M}
\left[1-{\rm exp}(-{\rm ln}2(\frac{\theta_{\rm s}}{\theta_{\rm M}})^{2})\right],
\end{equation}
\begin{equation}
\label{eqn:eff4}  
\Omega_{\rm M} = 1.133\,\theta^{2}_{\rm M},
\end{equation} 
where $A_{\rm e}$, $A_{\rm p}$, $\lambda$, $T_{\rm B}$, $\Omega_{\rm s}$
are the effective and physical areas of main reflector, 
the observing wavelength,
the brightness temperature, the source solid angle, and $\Omega_{\rm M}$,
$\Omega_{\rm A}$, $\theta_{\rm s}$, $\theta_{\rm M}$ are 
the main beam solid angle, the antenna solid angle, the angular size of 
the source, and the size of main beam (FWHM), respectively.

Table~\ref{tab-eff} lists measurements of main beam size, $\theta_{\rm M}$,
apperture efficiency, $\eta_{\rm A}$,
and main beam efficiency, $\eta_{\rm B}$
at 22 and 43\,GHz 
over the period 2009 April--2011 March.
The measured aperture efficiencies are 
$65\%$(K)/$67\%$(Q), 
$62\%$(K)/$59\%$(Q), 
and $66\%$(K)/$60\%$(Q) 
for KVNYS, KVNUS, and KVNTN, respectively.
The measurement errors of the aperture efficiencies are less than 2\%.
Main beam sizes were estimated from the CS or OTF observations
of Venus or Jupiter
as well as from the FIVE observations of strong \water and SiO maser sources. 
For each measurement of the apperture efficiency, 
Jupiter or Venus have been observed in CS or OTF mapping mode 
over various elevations when the angular sizes of the planets
were smaller than the main beam size at 43\,GHz. 
Measurements at elevations of $\geq 30^{\circ}$ 
were averaged to estimate the main beam size and the apperture efficiency
with their standard deviations listed in Table~\ref{tab-eff}.
The main beam efficiency was calculated through its relation 
to the apperture efficiency (Eq.[\ref{eqn:eff2}]).

The uncertainties of the brightness temperatures of 
Venus ($505\pm25$\,K at K-band~\citep{but+01} 
and $450\pm32$\,K at Q-band~\citep{gre+94}) 
and Jupiter ($134\pm4$\,K at K-band~\citep{pag+03} 
and $150\pm12$\,K at Q-band~\citep{gre+94})
dominate the error of the aperture efficiency.
The predicted aperture efficiencies from the original design
of the antenna with a blockage efficiency of 88\% are 
72\% at K band and 75\% at Q band.
The measured aperture efficiencies are lower than these 
by $\le$10\% and $\le$16\% at each band.
This is, in fact, expected from the results of the 100\,GHz 
test observations of the KVN radio telescopes~\citep{kim+11}.
The 100\,GHz aperture efficienies are 52\%
for the three antennas, which are lower than the designed value
of 60\%, due to a slightly lower blockage efficiency
from the additional cladding of the subreflector supports.
The aperture efficiencies at 100\,GHz are similar
for all antennas, whereas those at K and Q bands
are different from each antenna. At K band, the aperture efficiency
of KVNUS is lower than those of the others, and at Q band,
KVNUS and KVNTN have lower aperture efficincies than KVNYS.  
This may be 
because of larger misalignment of quasioptical system at KVNUS and KVNTN
than that at KVNYS.
The aperture efficiency is degraded if the quasioptical system
is not well aligned with the antenna,
although the relative beam alignment between two frequency bands is good.
The alignment will be improved when 
new receivers at 86 and 129\,GHz and corresponding quasioptical systems are
installed.

\begin{figure}
\epsscale{1.0}
\plotone{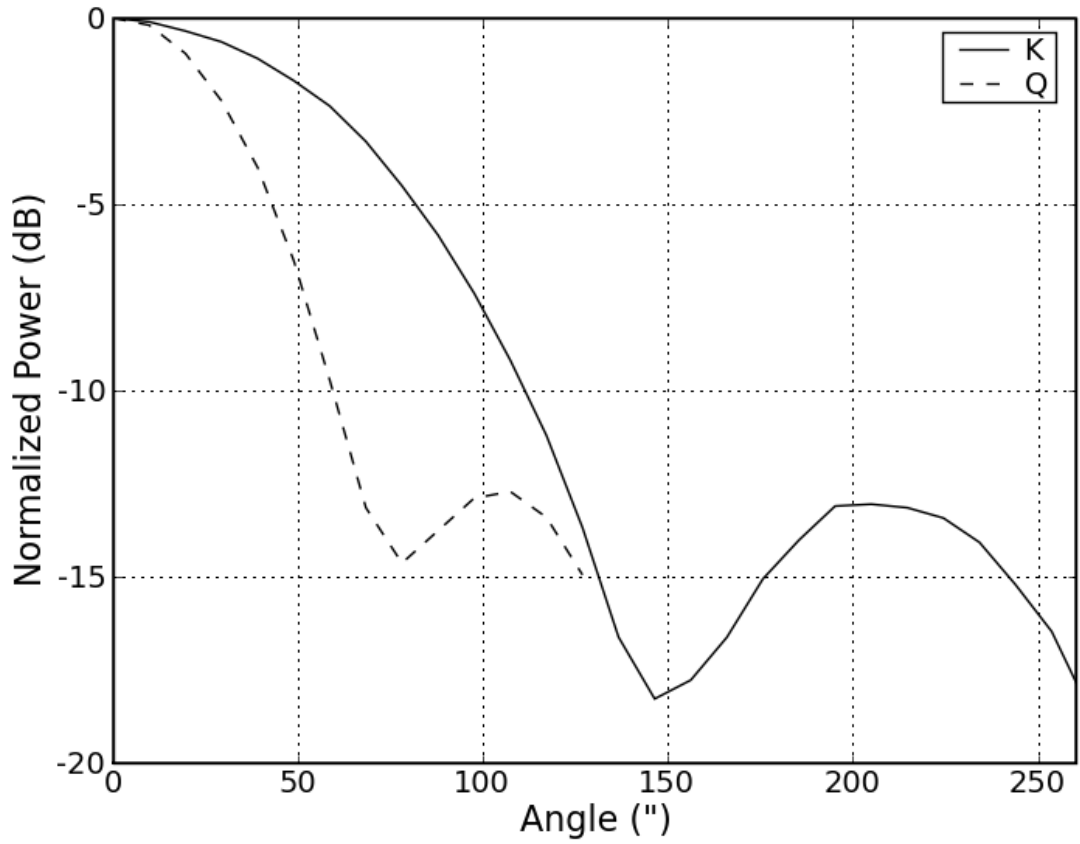}
\caption{Azimuthally averaged beam patterns at K-band (solid line) and
Q-band (dashed line) for KVNTN from OTF mapping observations of Jupiter.
\label{fig:beam}}
\end{figure}

The measured main beam sizes are
$122^{\prime\prime}$(K)/$64^{\prime\prime}$(Q), 
$123^{\prime\prime}$(K)/$66^{\prime\prime}$(Q), 
and
$124^{\prime\prime}$(K)/$64^{\prime\prime}$(Q) 
for KVNYS, KVNUS, and KVNTN, respectively over the period
2009 April--2011 March (Table~\ref{tab-eff}).
The main beam sizes have no dependence on elevation and 
a small difference of $<$4\% between azimuth and elevation directions.
These main beam sizes are smaller by $\sim10\%$ than the diffraction limit
of $135^{\prime\prime}$ at 22.235\,GHz and 
$70^{\prime\prime}$ at 43.122\,GHz for uniform aperture illumination. 
This may result from blockage effect~\citep{hei+01} and/or
inverse-tapered illumination of the main reflector.
These effects also result in
relatively higher sidelobe levels than for the conventional Cassegrain antenna.
The antenna beam patterns for all KVN antennas were measured
and the beam pattern for KVNTN is shown in Figure~\ref{fig:beam}. 
The first sidelobe levels are 
-13.0/-13.0\,dB (K/Q), -13.0/-12.7\,dB (K/Q), and -13.1/-12.7\,dB (K/Q)
for KVNYS, KVNUS, and KVNTN, respectively.
The measured sidelobe levels are significantly higher than those (-17.6\,dB) for 
the Cassegrain antennas with uniform illumination pattern 
and without blockage~\citep{rw00}. 
It should be noticed
that, due to the higher sidelobe level, 
spectral line observation data 
should be carefully analyzed, taking into account
possible coupling to the first sidelobe.

The main-beam efficiencies are measured to be 
$50\%$(K)/$52\%$(Q), 
$48\%$(K)/$50\%$(Q), 
and $50\%$(K)/$47\%$(Q) 
for KVNYS, KVNUS, and KVNTN, respectively.
Since the main-beam efficiency was estimated according to its relation 
to the apperture efficiency (Eq.[\ref{eqn:eff2}]),
the statistical errors are not estimated.
However the estimation of the main-beam efficiencies are also affected
by the uncertainty of the brightness temperatures of Venus and Jupiter
and the measurement errors of main-beam sizes.
Due to the high level sidelobes, the main-beam efficiencies are
lower with respect to the aperture efficiencies than in case of 
conventional Cassegrain telescopes.

We measured Moon efficiencies, $\eta_{\rm moon}$,
of the KVN radio telescopes by OTF-mapping the Moon
at 22 and 43\,GHz. The Moon efficiencies  
were derived based on \cite{lin73} and \cite{man93}.
The estimated Moon efficiencies are 
$77\%$(K)/$90\%$(Q), 
$74\%$(K)/$79\%$(Q),
and $80\%$(K)/$86\%$(Q) 
for KVNYS, KVNUS, and KVNTN, respectively (Table~\ref{tab-meff}).
The error of the Moon efficiency is dominated by 
the measurement error of antenna temperature of Moon which is
as large as 1-2\%.

\subsubsection{Gain curve}
Gain curve measurements were made
in September, 2009 for KVN Yonsei,
in March, 2011 for KVN Ulsan,
and in December, 2010 for KVN Tamna.
The elevation dependence of the aperture efficiency of
the KVN 21-m radio telescopes
was measured by observing bright \water and SiO maser sources 
at 22 and 43\,GHz in dual polarizations (LCP and RCP) 
using the FIVE pointing observations.
Every 10 FIVE pointing observations, 
the atmospheric opacity was measured and the subreflector focus was adjusted
based on the pre-established model for the subreflector.
In order to avoid gain degradation due to the thermal deformation
of antenna, the gain curve measurements were performed at night.

\begin{figure}
\epsscale{1.0}
\plotone{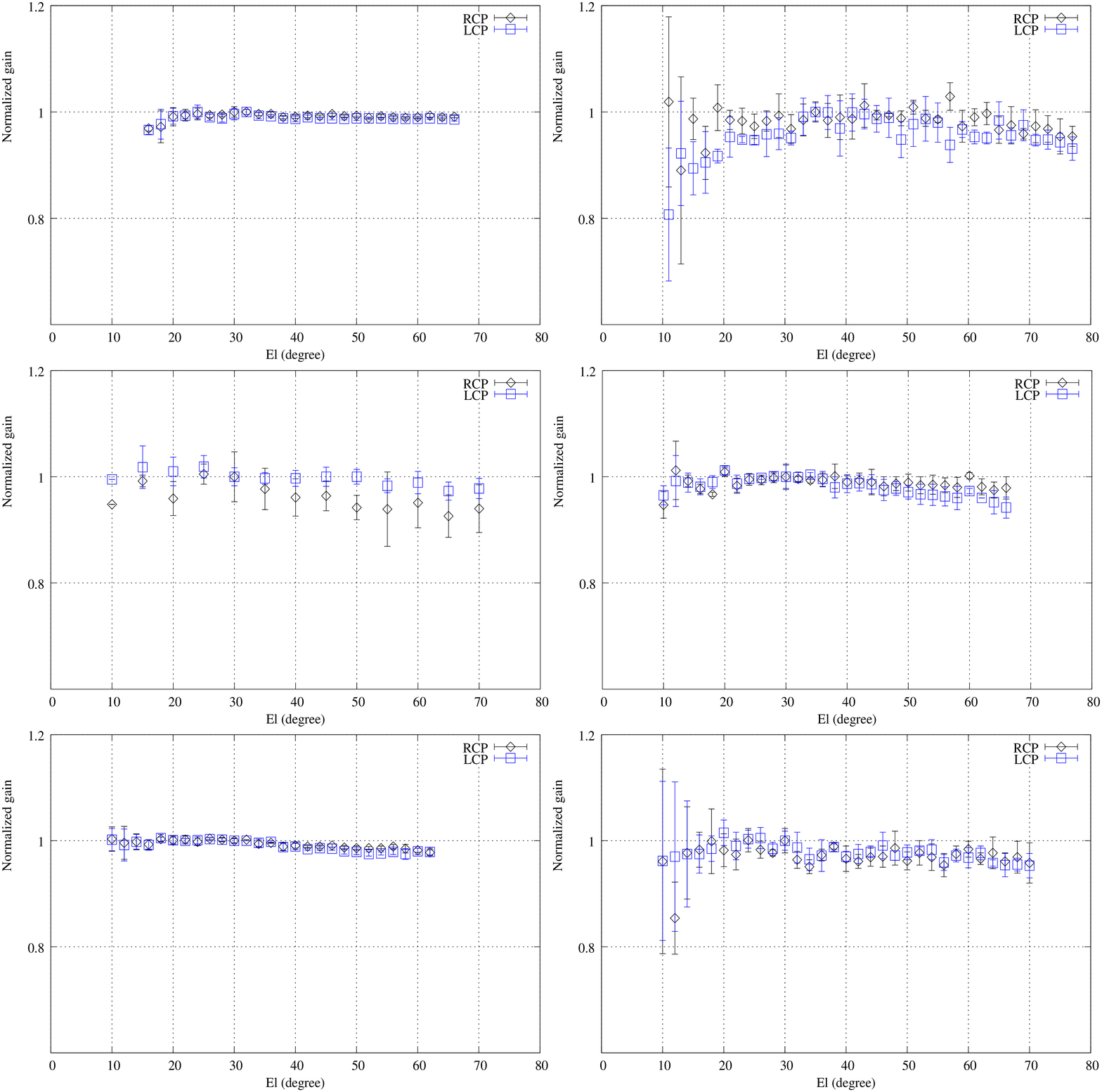}
\caption{Gain curves of KVN Yonsei (upper), Ulsan (middle), Tamna (bottom)
at 22\,GHz (left) and 43\,GHz (right).
\label{fig:gc}}
\end{figure}

The pointing offsets were corrected during the data reduction based on 
the results of the FIVE pointing observations. 
Figure~\ref{fig:gc} shows 
the normalized specral line intensities of \water and SiO maser sources
in elevations of $10^{\circ}-70^{\circ}$
for KVN Yonsei, Ulsan, and Tamna at 22 and 43\,GHz.
A second order polynomial was least-square-fitted to the data.
The fitted function was normalized by its maximum.
Normalized gain curve has the following form:
${\rm Gain}_{\rm norm} = {\rm A0}\times El^2+{\rm A1}\times El+{\rm A2}$,
where $El$ is the elevation in degree.
The fitted parameters are summarized in Table~\ref{tab-gc}.
The gain curves at 22\,GHz
are quite flat at the elevatations of $>20^{\circ}$. 
At 43\,GHz,
the antenna gains 
at low elevation decrease by less than 10\%. 

\section{Some astronomical results}
\subsection{Maser line surveys of the Galactic star-forming regions}

Extensive simultaneous H$_2$O $6_{16}-5_{23}$ (22.23508~GHz) and 
CH$_3$OH $7_{0}-6_{1}~A^{+}$ (44.06943~GHz) maser line surveys 
were carried out towards Galactic young stellar objects (YSOs) 
using the KVN 21-m telescopes.
Because of variability of the two masers,
simultaneous observations are important
for comparison of their properties in detail.
\cite{bae+11} report the observational results on 180 intermediate-mass 
YSOs, including
14 Class~0 objects, 19 Class~I objects, and 147 Herbig Ae/Be stars that
are widely believed to be intermediate-mass stars 
in the pre-main sequence phase.
\water\ and \methanol\ maser emission were detected towards 16 (9\%) 
and 10 (6\%) sources, respectively. The detection rates 
of both masers rapidly decrease as the central (proto)stars evolve,
which is consistent with the tend for H$_2$O masers 
observed in low-mass star-forming regions~\citep{fur+01}.
This indicates that the excitation of the two masers is related to
the evolutionary stage of the central object.
\water\ masers usually have significantly larger relative velocities
with respective to the ambient molecular gas than do \methanol\ masers. 
The isotropic luminosities of both masers are well 
correlated with the bolometric luminosities of the central (proto)stars.

The same maser line surveys were performed towards more than 1000 high-mass YSOs
in different evolutionary phases: infrared dark cores, high-mass
protostellar candidates, and ultracompact HII regions 
(Kim et al. in preparation). 
The detection rates of both masers significantly increase as the central
objects evolve. 
This is contrary to the trends found in low- and intermediate-mass 
star-forming regions, as mentioned above. 
Thus the excitation 
of the two masers appear to be closely related to the circumstellar environments
as well as the evolutionary stage of the central (proto)stars.
\subsection{Late-type stars}
\begin{figure}
\epsscale{1.0}
\plotone{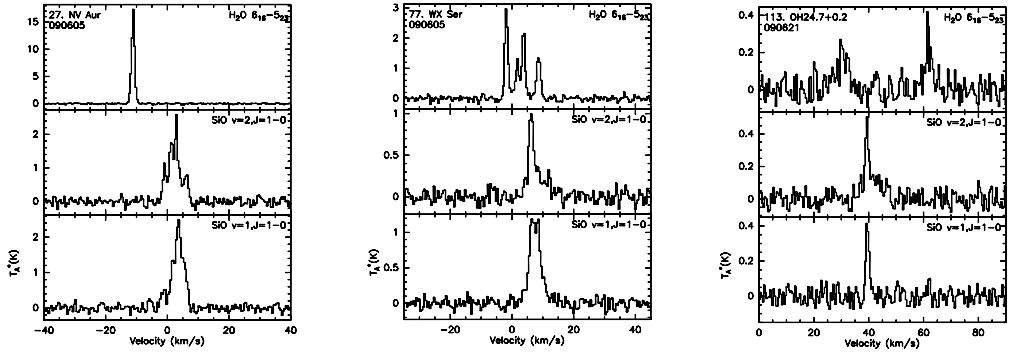}
\caption{Sample spectra of three late-type stars (NV Aur, WX Ser, and OH24.7+0.2) 
simultaneously obtained in the \water $6_{16}-5_{23}$ and SiO $v=1$ and $2, J=1-0$ 
transitions. Adopted from \cite{kim+10}.
}
\label{fig:ls}
\end{figure}

Studies of late-type stars using the KVN Yonsei radio telescope 
were performed toward known stellar SiO and/or \water maser sources 
(166 both SiO and \water maser sources, 83 SiO-only sources, 
and 152 ${\rm H}_{2}{\rm O}$-only sources) 
soon after finishing the performance test 
observations. Both SiO and \water masers were detected from 112 sources
at one epoch giving a detection rate of 67\% toward 166 both
SiO and \water maser sources~\citep{kim+10},
detected from 14 sources (detection rate of 17\%) 
toward 83 SiO-only sources (Cho et al. 2011 in prep.),
and detected from 62 sources (detection rate of 41\%)
toward 152 ${\rm H}_{2}{\rm O}$-only sources (Kim et al. 2011 in prep.), 
respectively.
Some spectra simultaneously obtained in 
the \water\,$6_{16}-5_{23}$ and SiO $v=1$ and 2, $J=1-0$ transitions
are shown in Figure~\ref{fig:ls}.
Through simultaneous observations of SiO and \water masers, 
relations between SiO and \water maser properties 
and the dynamical connection from the pulsating atmosphere 
to the inner circumstellar envelope were investigated 
toward evolved stars. 
The distribution of single and double peaked profiles 
of \water maser lines were also examined in the IRAS two-color diagram. 
These single and double peaked \water maser lines 
can be associated with an asymmetric wind and bipolar outflows 
which are commonly seen in proto-planetary nebulae and 
planetary nebulae as suggested by \cite{eng02}.

\subsection{Intraday variable source 0716+714}

\begin{figure}
\epsscale{.96}
\plotone{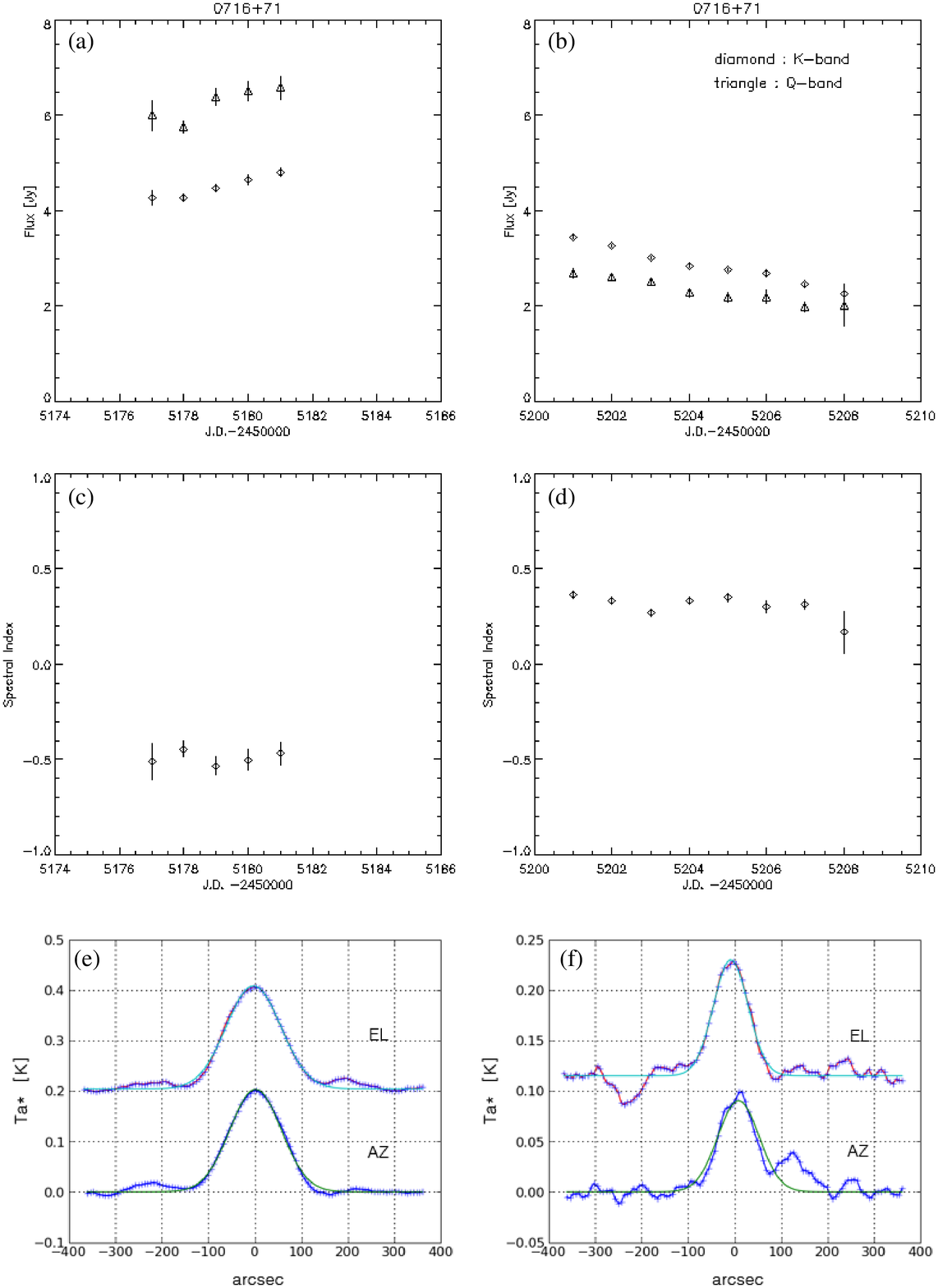}
\caption{ 
Light curves of 0716+714 
in December 2009(JD 24505177-24505181) (a) and 
in January 2010(JD 24505201-24505208) (b), respectively,
and spectral indices for the measurements in December 2009 (c) and
in January 2010 (d), repectively. 
Examples of cross scan profiles with their gaussian fits
of a flux calibrator 3C 286 taken simultaneously
at 21.7\,GHz (e) and 42.4\,GHz (f).
Error bars (one-sigma) in (a)-(d) are statistical only.
Effective on-source integration times
for the examples (e) and (f) are 40 and 20 seconds, respectively.
}
\label{fig:0716}
\end{figure}

Total flux monitoring of the intraday variable source 0716+714 were made
simultaneously at 21.7\,GHz and 42.4\,GHz with KVN Yonsei 21-m 
radio telescope on December 11-15, 2009 and January 4-11, 2010. 
An absolute calibrator, 3C 286, was observed for flux calibration,
and a nearby secondary calibrator, 0836+714, observed 
for gain calibration.
All flux measurements were done with CS (Cross Scan) mode. 
The total bandwidth was 500\,MHz at both frequency bands.
Figure~\ref{fig:0716} shows the light curves and specral indicies
of 0716+714 for two epochs.
In these observations, 
intraday variability of 0716+714 is not seen, 
but a monotonic increase and decrease of total flux density 
on longer time scale were observed. 
More interestingly, as the source gets brighter 
the spectral index becomes flatter,
and vice-versa.

\section{Summary}

Through the performance evaluation observations in 2008-2011, we found that
the KVN 21-m radio telescopes are suitable for simultaneous
multi-frequency single polarization observations at 22 and 43\,GHz.
The dual beams at two frequencies are well aligned within $<5$ arcseconds.
The pointing accuracies are $<$3 arcseconds in azimuth and elevation. 
The measured aperture efficiencies are of $>64\%$ at 22\,GHz 
and $>62\%$ at 43\,GHz, 
the main-beam efficiencies are $>49\%$ at 22\,GHz and $>50\%$ at 43\,GHz, and
the estimated Moon efficiencies are $>77$ at 22\,GHz and $>85\%$ at 43\,GHz.
The elevation dependence of the aperture efficiencies is quite flat at
the elevations of $> 20^{\circ}$
Unique receiving systems, accurate beam alignment, and high
antenna efficiencies will provide us with good opportunities for
scientific observations at millimeter wavelengths not only for
single radio telescope but also for Very Long Baseline Interferometry.

Further instrumentaion for KVN will include 86 and 129\,GHz receivers
which should be ready in late 2012. Simultaneous four-frequency observations
will be ready for VLBI observations.

\acknowledgments

This work was supported by Basic Research Program (2008-2011) and
also partially supported by KASI-Yonsei Joint Research Program (2010-2011)
for the Frontiers of Astronomy and Space Science funded by
the Korea Astronomy and Space Science Institute.
We are grateful to Paul Goldsmith for careful reading and kind comments 
to the manuscript. We would like to thank the anonymous referee for
providing prompt and thoughtful comments that helped improve
the original manuscript.


\clearpage

\begin{deluxetable}{ll}
\tabletypesize{\scriptsize}
\tablecaption{Specification of KVN antennas\label{tbl-1}}
\tablewidth{0pt}
\tablehead{
Reflector characteristics
}
\startdata
Main Reflector (axisymmetric paraboloid)& \\
\hspace{0.5cm} diameter & D=21.03\,m \\
\hspace{0.5cm} focal length & f=6.78\,m \\
\hspace{0.5cm} focal ratio & f/D=0.32 \\
\hspace{0.5cm} panels manufacturing accuracy & 65\,$\mu m$ \\
\hspace{0.5cm} alignment surface accuracy & 50-54\,$\mu m$ \\
Subreflector (hyperboloid) & \\
\hspace{0.5cm} diameter & d=2.25\,m\\
\hspace{0.5cm} manufacturing surface accuracy & 50\,$\mu$m\\
Expected total surface ${\rm accuracy} ({\rm EL}\sim48^{\circ})^{\dagger}$ & 124\,$\mu$m\\
\hspace{0.5cm} panel & 73\,$\mu$m\\
\hspace{0.5cm} alignment & 60\,$\mu$m\\
\hspace{0.5cm} subreflector & 52\,$\mu$m\\
\hspace{0.5cm} backup structure & 62\,$\mu$m\\
\tableline
Environmental constraints & \\
\tableline
\hspace{0.5cm} slewing speed  & $3^{\circ}\,\rm{sec}^{-1}$ \\
\hspace{0.5cm} slewing acceleration  & $3^{\circ}\,\rm{sec}^{-2}$ \\
\hspace{0.5cm} operating range  & AZ:$\pm 270^{\circ}$ \\
\hspace{0.5cm}                  & EL:$0^{\circ}-90^{\circ}$ \\
\enddata
\tablecomments{$\dagger$:Aggregate surface accuracy of KVN antennas
obtained from RSS (Root-Sum-Square) calculation of
measured and expected contributions 
by main reflector panels, alignment,
subrefelector, and antenna backup structures.
Individual contribution includes measured accuracies, measurement errors,
expected thermal and wind effects.
}
\end{deluxetable}

\clearpage
\begin{deluxetable}{ccccccccc}
\tabletypesize{\scriptsize}
\tablecaption{Singal losses by dichroic filters \label{tbl-lpf}}
\tablewidth{0pt}
\tablehead{
\colhead{} & 
\multicolumn{2}{c}{LPF1} &
\colhead{} & 
\multicolumn{2}{c}{LPF2} & 
\colhead{} & 
\multicolumn{2}{c}{LPF1+LPF2} \\
\cline{2-3}\cline{5-6}\cline{8-9}\\
\colhead{Center freq.} & 
\colhead{LCP} & \colhead{RCP} & 
\colhead{} & 
\colhead{LCP} & \colhead{RCP} & 
\colhead{} & 
\colhead{LCP} & \colhead{RCP}\\
\colhead{(GHz)} & 
\colhead{(\%)} & \colhead{(\%)} & 
\colhead{} & 
\colhead{(\%)} & \colhead{(\%)} & 
\colhead{} & 
\colhead{(\%)} & \colhead{(\%)}
}
\startdata
21.50&0.70&	1.16&&	2.48&	3.11&&	3.15&	4.20 \\
21.75&0.84&	1.03&&	2.89&	2.84&&	3.69&	3.82 \\
22.00&0.80&	0.80&&	2.53&	2.57&&	3.30&	3.32 \\
22.25&0.77&	0.70&&	2.52&	2.39&&	3.25&	3.06 \\
22.50&0.58&	0.65&&	2.52&	2.43&&	3.07&	3.05 \\
22.75&0.77&	0.39&&	2.05&	2.64&&	2.78&	3.01 \\
23.00&0.47&	0.46&&	2.63&	2.57&&	3.07&	3.01 \\
\hline
42.36&1.08&	1.08&&	3.02&	5.38&&	4.03&	6.35 \\
42.61&0.75&	1.47&&	5.30&	5.59&&	5.97&	6.91 \\
42.86&1.06&	1.00&&	3.99&	4.26&&	4.97&	5.17 \\
43.11&1.29&	0.88&&	3.01&	2.90&&	4.23&	3.74 \\
43.36&1.34&	1.21&&	2.30&	2.35&&	3.57&	3.50 \\
43.61&1.30&	1.07&&	1.67&	1.90&&	2.93&	2.93 \\
43.86&1.48&	1.15&&	1.82&	1.82&&	3.24&	2.93 \\
\enddata
\tablecomments{Frequency bandwidth for each measurement is 500\,MHz.
The measurements were done on 17 July, 2008. 
}
\end{deluxetable}

\clearpage
\begin{deluxetable}{lccc} 
\tabletypesize{\scriptsize}
\tablecolumns{4} 
\tablewidth{0pt} 
\tablecaption{The Available DSM Ouputs}
\tablehead{ 
\colhead{} & \colhead{} & \colhead{Bandwidth\tablenotemark{a}} & \colhead{No. of} \\
\colhead{Mode} & \colhead{} & \colhead{(MHz)} & \colhead{Streams}
} 
\startdata 
Wide   & & 512 & 4  \\
Narrow &1 & 256 & 1  \\
&2 & 128 & 2  \\
&3 & 64 & 4  \\
&4 & 32 & 8  \\
&5 & 16 & 8  \\
&6 & 8 & 8  \\
&7 & 64/128& 2/1  \\
&8 & 32/64/128& 2/1/1  \\
&9 & 32/128& 4/1  \\
&10 & 16/32/128& 2/3/1  \\
\enddata 
\tablecomments{ 
Half bandwidth can be taken with the same number of channels for all outputs.
}
\label{table:obsmode}
\end{deluxetable} 

\clearpage

\begin{deluxetable}{lllccc}
\tabletypesize{\scriptsize}
\tablecaption{Measured beam alignments of KVN antennas at 22 and 43\,GHz\label{tab-bl}}
\tablewidth{0pt}
\tablehead{
\colhead{} & \colhead{} & \colhead{$\Delta_{\rm Az}$} &
\colhead{$\Delta_{\rm El}$} &\colhead{} &\colhead{}\\
\colhead{Telescope} & \colhead{Epoch} & \colhead{($\pp$)} &
\colhead{($\pp$)} &\colhead{Source} &\colhead{Obs. mode}}
\startdata
KVNYS    &  Sep. 2009 & $-$2.3 & $+$5.6 & VX Sgr & FIVE\\
         &            & $-$0.5 & $+$4.2 & Venus  & CS\\
         &            & $-$1.4 & $+$4.6 & Jupiter & CS\\
\cline{1-6}
KVNUS    &  Sep. 2009 & $-$4.2 & $+$3.3 & VX Sgr & FIVE\\
         &            & $-$0.7 & $+$3.3 & Venus  & CS\\
         &            & $-$3.2 & $+$2.9 & Jupiter & CS\\
\cline{1-6}
KVNTN    &  Dec. 2009 & $+$1.3 & $-$1.3 & VX Sgr & FIVE\\
         &            & $+$3.4 & $-$4.3 & Mars   & CS\\
         &            & $+$2.4 & $-$1.8 & Jupiter & CS\\
\enddata
\tablecomments{Offsets of the beam at 22\,GHz measured with respect to the beam at 43\,GHz.
}
\end{deluxetable}

\clearpage

\begin{deluxetable}{lllc}
\tabletypesize{\scriptsize}
\tablewidth{0pt}
\tablecaption{SiO maser sources used for the pointing model\label{tab-poi1}}
\tablehead{
\colhead{} & \colhead{$\alpha_{2000}$} & \colhead{$\delta_{2000}$} &
\colhead{$V_{\rm LSR}$}\\
\colhead{Source} & \colhead{(hh mm ss)} & \colhead{(dd mm ss)} &
\colhead{(km/s)}}
\startdata
R Cas    &  23 58 24.79 & $+$51 23 19.5 &  $+$25.0\\
U Her    &  16 25 47.48 & $+$18 53 33.0 &  $-$15.0\\
R Leo    &  09 47 33.49 & $+$11 25 44.0 &  $-$1.0\\
IK Tau   &  03 53 28.87 & $+$11 24 21.7 &  $+$37.1\\
{\it o} Cet &  02 19 20.79 &$-$02 58 37.4 &  $+$50.2\\
R Aqr    &  23 43 49.44 & $-$15 17 03.9 &  $+$22.0\\
VX Sgr   &  18 08 04.05 & $-$22 13 26.6 &  $+$5.7\\
W Hya    &  13 49 02.03 & $-$28 22 03.0 &  $+$42.0\\
VY CMa   &  07 22 58.33 & $-$25 46 03.2 &  $+$18.0\\
\enddata
\end{deluxetable}

\clearpage

\begin{deluxetable}{lcccc}
\tabletypesize{\scriptsize}
\tablecaption{Pointing accuracies of KVN antennas at 43GHz\label{tab-poi2}}
\tablewidth{0pt}
\tablehead{
\colhead{} & \colhead{} & \colhead{$\sigma_{\rm Az}$} &
\colhead{$\sigma_{\rm El}$} & \colhead{$\sigma_{\rm tot}$}\\
\colhead{Telescope} & \colhead{Epoch} & \colhead{($\pp$)} &
\colhead{($\pp$)} & \colhead{($\pp$)}}
\startdata
KVNYS    &  6 Mar. 2009 & 2.2 & 4.0  & 4.6 \\
         &  18 Sep. 2009 & 2.2 & 5.0 & 5.5 \\
         &  16 Oct. $2010^{\dagger}$ & 1.7 & 2.6 & 3.1  \\
\cline{1-5}
KVNUS    &  9 Sep. 2009 & 2.9 & 4.9 & 5.7\\
         &  8 Oct. 2010 & 2.2 & 5.5 & 5.9  \\
         &  8 Nov. $2010^{\dagger}$ & 2.8 & 2.3 & 3.6  \\
\cline{1-5}
KVNTN    &  7 Dec. 2009 & 2.4 & 5.3 & 5.8\\
         &  17 Sep. 2010 & 2.6 & 5.7 & 6.3  \\
         &  10 Jan. $2011^{\dagger}$ & 2.0 & 2.8 & 3.4  \\
\enddata
\tablecomments{$\dagger$: Pointing models established after improving
the pointing accuracies by applying a cladding system.
}
\end{deluxetable}

\clearpage

\begin{deluxetable}{lcrrccccc}
\tabletypesize{\scriptsize}
\tablecaption{Efficiencies of KVN antennas at 22 and 43\,GHz
\label{tab-eff}}
\tablewidth{0pt}
\tablehead{
\colhead{} &\colhead{} & 
\colhead{} &\colhead{$\theta_{\rm M}$} &
\colhead{$\eta_{\rm A}$} & \colhead{$\eta_{\rm B}$} &
\colhead{} &
\colhead{Ang. size}& \colhead{Elevation} \\
\colhead{Telescope} &\colhead{Band} & 
\colhead{Epoch}     &\colhead{$(\pp)$} &
\colhead{$(\%)$}    &\colhead{$(\%)$} &
\colhead{Target} &
\colhead{($\prime\prime$)}& \colhead{($\circ$)}} 
\startdata
KVNYS    &  K &7 Apr. 2009 & $124.6\pm0.6$ & $71.9\pm1.2$& 56.6 & Venus&56.1  &30-59\\
        &  Q &8 Apr. 2009 & $65.1\pm2.4$ & $68.4\pm0.8$& 55.3 & Venus&55.5  &30-59\\
        &  K &10 Nov. 2010 & $123.4\pm1.2$ & $61.4\pm0.5$& 47.4 & Jupiter&45.8  &30-49\\
        &  Q &11 Nov. 2010 & $66.8\pm0.6$ & $64.3\pm0.4$& 54.8 & Jupiter&45.7  &30-49\\
        &  K &16 Mar. 2011 & $120.9\pm1.1$ & $66.6\pm0.8$& 49.4 & Venus&14.4  &30-34\\
        &  Q &16 Mar. 2011 & $62.1\pm0.6$ & $67.3\pm1.0$& 49.5 & Venus&14.4  &30-34\\
        &  K &16 Mar. 2011 & $120.5\pm1.7$ & $61.5\pm0.8$& 45.3 & Jupiter&33.4  &30-56\\
        &  Q &16 Mar. 2011 & $61.5\pm0.9$ & $67.7\pm1.0$& 48.9 & Jupiter&33.4  &30-56\\
\cline{1-9}
KVNUS   &  K &17 Sep. 2009 & $124.4\pm3.3$ & $64.5\pm1.9$& 50.6 & Venus&11.8  &30-59\\
        &  Q &18 Sep. 2009 & $65.9\pm3.9$ & $67.6\pm0.7$& 56.0 & Venus&11.7  &30-59\\
	&  K &10 Nov. 2010 & $124.0\pm0.3$ & $60.2\pm0.5$& 47.0 & Jupiter&45.8 &30-50\\
        &  Q &10 Nov. 2010 & $68.2\pm0.2$ & $56.3\pm0.9$& 50.0 & Jupiter&45.8 &30-50\\
	&  K &16 Mar. 2011 & $121.5\pm0.9$ & $64.6\pm0.6$& 48.4 & Venus&14.4 &30-39\\
        &  Q &16 Mar. 2011 & $64.3\pm0.7$ & $57.1\pm0.8$& 45.0 & Venus&14.4 &30-39\\
	&  K &16 Mar. 2011 & $122.9\pm1.0$ & $60.1\pm0.3$& 46.1 & Jupiter&33.4 &30-58\\
        &  Q &16 Mar. 2011 & $66.1\pm0.6$ & $56.9\pm1.1$& 47.4  & Jupiter&33.4 &45-53\\
\cline{1-9}
KVNTN   &  K &15 Feb. 2011 & $121.4\pm0.8$ & $69.4\pm0.4$ &51.9& Venus&17.5  &30-35\\
        &  Q &15 Feb. 2011 & $63.3\pm0.3$ & $60.3\pm0.4$ &46.1&  Venus&17.5  &30-35\\
        &  K &15 Feb. 2011 & $122.8\pm0.6$ & $63.4\pm0.4$ &48.5&  Jupiter&34.7  &30-57\\
        &  Q &15 Feb. 2011 & $65.3\pm0.4$ & $59.5\pm1.4$ &48.4&  Jupiter&34.7  &30-57\\
\enddata
\tablecomments{ The brightness temperatures used of Jupiter/Venus are
$134\pm4$\,K~\citep{pag+03}/$505\pm25$\,K~\citep{but+01} at K-band and 
$150\pm12$\,K~\citep{gre+94}/$450\pm32$\,K~\citep{gre+94} at Q-band. 
The uncertainties of the main beam size and the aperture efficiency
include only statistical measurement errors.
}
\end{deluxetable}

\clearpage

\begin{deluxetable}{lcrcccc}
\tabletypesize{\scriptsize}
\tablecaption{Moon efficiencies of KVN antennas at 22 and 43\,GHz
\label{tab-meff}}
\tablewidth{0pt}
\tablehead{
\colhead{} &
\colhead{} & 
\colhead{} & 
\colhead{UT} &\colhead{phase} &
\colhead{$T_{\rm a}^{*}$} &
\colhead{$\eta_{\rm moon}$}\\
\colhead{Telescope} &
\colhead{Band}   &
\colhead{Epoch} & 
\colhead{(hh:mm)}     &\colhead{(days)} &
\colhead{(K)}    &\colhead{$(\%)$} 
} 
\startdata
KVNYS   &K&  15 Mar. 2011 & 18:30 & 10.5 & 184.1 & 76.9\\
	&Q&  15 Mar. 2011 & 18:30 & 10.5 & 215.0 & 89.6\\
KVNUS   &K&  16 Mar. 2011 & 19:20 & 11.5 & 181.7 & 73.6\\
	&Q&  16 Mar. 2011 & 19:20 & 11.5 & 198.8 & 78.8\\
KVNTN   &K&  15 Feb. 2011 & 12:50 & 12.4 & 202.0 & 79.8\\
	&Q&  15 Feb. 2011 & 12:50 & 12.4 & 224.6 & 85.7\\
\enddata
\tablecomments{ 
We used 239\,K and 234\,K at K and Q bands for the brightness temperatures 
of Moon based on \cite{lin73}. 
}
\end{deluxetable}

\clearpage

\begin{deluxetable}{lcrccc}
\tabletypesize{\scriptsize}
\tablecaption{Normalized gain curve parameters of KVN antennas at 22 and 43\,GHz
\label{tab-gc}}
\tablewidth{0pt}
\tablehead{
\colhead{Telescope} &
\colhead{Band}   &
\colhead{Epoch} & 
\colhead{A0}     &
\colhead{A1} &
\colhead{A2} 
} 
\startdata
KVNYS   &K&  Sep. 2009 & $-6.70130\times 10^{-6}$ & $+4.85791\times 10^{-4}$ & 0.991196 \\ 
        &Q&  Sep. 2009 & $-3.94673\times 10^{-5}$ & $+3.47852\times 10^{-3}$ & 0.923354\\ 
KVNUS   &K&  Mar. 2011 & $-1.39267\times 10^{-6}$ & $-5.56796\times 10^{-4}$ & 1.005707\\ 
        &Q&  Mar. 2011 & $-1.94371\times 10^{-5}$ & $+1.41639\times 10^{-3}$ & 0.974197\\ 
KVNTN   &K&  Dec. 2009 & $-1.35247\times 10^{-5}$ & $+6.81293\times 10^{-4}$ & 0.991420\\ 
        &Q&  Dec. 2009 & $-1.91393\times 10^{-5}$ & $+1.23772\times 10^{-3}$ & 0.979990\\ 

\enddata
\end{deluxetable}

\clearpage

\clearpage

\clearpage

\clearpage

\clearpage

\clearpage

\clearpage

\clearpage

\clearpage

\clearpage

\end{document}